\documentclass[sigconf]{acmart}

\AtBeginDocument{%
  \providecommand\BibTeX{{%
    \normalfont B\kern-0.5em{\scshape i\kern-0.25em b}\kern-0.8em\TeX}}}

\copyrightyear{2024}
\acmYear{2024}
\setcopyright{rightsretained}
\acmConference[Websci '24]{ACM Web Science Conference}{May 21--24, 2024}{Stuttgart, Germany}
\acmBooktitle{ACM Web Science Conference (Websci '24), May 21--24, 2024, Stuttgart, Germany}\acmDOI{10.1145/3614419.3644015}
\acmISBN{979-8-4007-0334-8/24/05}

\begin{document}

\title{Unveiling News Publishers Trustworthiness Through Social Interactions}


\author{Manuel Pratelli}
\email{manuel.pratelli@imtlucca.it}
\orcid{0000-0002-9978-791X}
\affiliation{
  \institution{IMT School for Advanced Studies Lucca\\ Institute of
Informatics and Telematics, CNR, Pisa}
 \country{Italy}
}

\author{Fabio Saracco}
\email{fabio.saracco@cref.it}
\orcid{0000-0003-0812-5927}
\affiliation{
  \institution{`Enrico Fermi' Research Center (CREF), Rome}
    \institution{IMT School for Advanced Studies Lucca}
\institution{
  Institute of Applied Mathematics ``Mauro Picone'' (IAC), CNR, Rome}
  \country{Italy}
  }

 

\author{Marinella Petrocchi}
\email{marinella.petrocchi@iit.cnr.it}
\orcid{0000-0003-0591-877X}
\affiliation{
  \institution{Institute of
Informatics and Telematics, CNR, Pisa \\ IMT School for Advanced Studies Lucca}
 \country{Italy}
}

\renewcommand{\shortauthors}{}

\begin{abstract}
With the primary goal of raising readers' awareness of misinformation phenomena, extensive efforts have been made by both academic institutions and independent organizations to develop methodologies for assessing the trustworthiness of online news publishers. Unfortunately, existing approaches are costly and face critical scalability challenges.
This study presents a novel framework for assessing the trustworthiness of online news publishers using user interactions on social media platforms. The proposed methodology provides a versatile solution that serves the dual purpose of i) identifying verifiable online publishers and ii) automatically performing an initial estimation of the trustworthiness of previously unclassified online news outlets. 
\end{abstract}

\ccsdesc[500]{Information systems~Decision support systems}
\ccsdesc[500]{Computing methodologies~Machine learning}

\keywords{Online Publishers Trustworthiness, Social Network Analysis, Community Detection, Information Disorder, Publishers/Consumers Interactions}



\maketitle

\section{Introduction}

The erosion of the mainstream journalism system, coupled with challenges in editorial control, has raised concerns about the quality of information disseminated by online media. Existing ratings by journalistic organizations and indexes, such as NewsGuard, MediaBias Fact Check, Iffy Index, Global Disinformation Index, and Ad Fontes Media\footnote{NewsGuard: https://www.newsguardtech.com; MediaBias Fact Check: https://mediabiasfactcheck.com/; Iffy Index: https://iffy.news/index/; The Global Disinformation Index: https://www.disinformationindex.org/; Ad Fontes Media: https://adfontesmedia.com/ - all URLs accessed on February 27, 2024.}, have and continue to shed light on the level of trustworthiness of many online news outlets. These ratings often take into account factors such as the tendency to publish propagandistic and/or politically biased content~\cite{bazmi2023multi,kim2019combating}. 

While these scores provide valuable insights, the labor-intensive nature of scoring individual news outlets is a significant time commitment. Manual analysis, in which expert annotators review ownership details and content, is essential but time-consuming~\cite{doi:10.1073/pnas.1806781116}.

In response, our current research seeks to streamline the evaluation process by leveraging social media interactions. Recognizing the central role of online platforms in shaping public discourse, our approach aims to circumvent the challenges posed by traditional evaluations. By examining the dynamics of social media interactions surrounding news publishers, we aim to provide a nuanced understanding of trustworthiness in the digital information landscape.

\begin{figure*}[h]
        \centering
        \includegraphics[width=.7\linewidth]{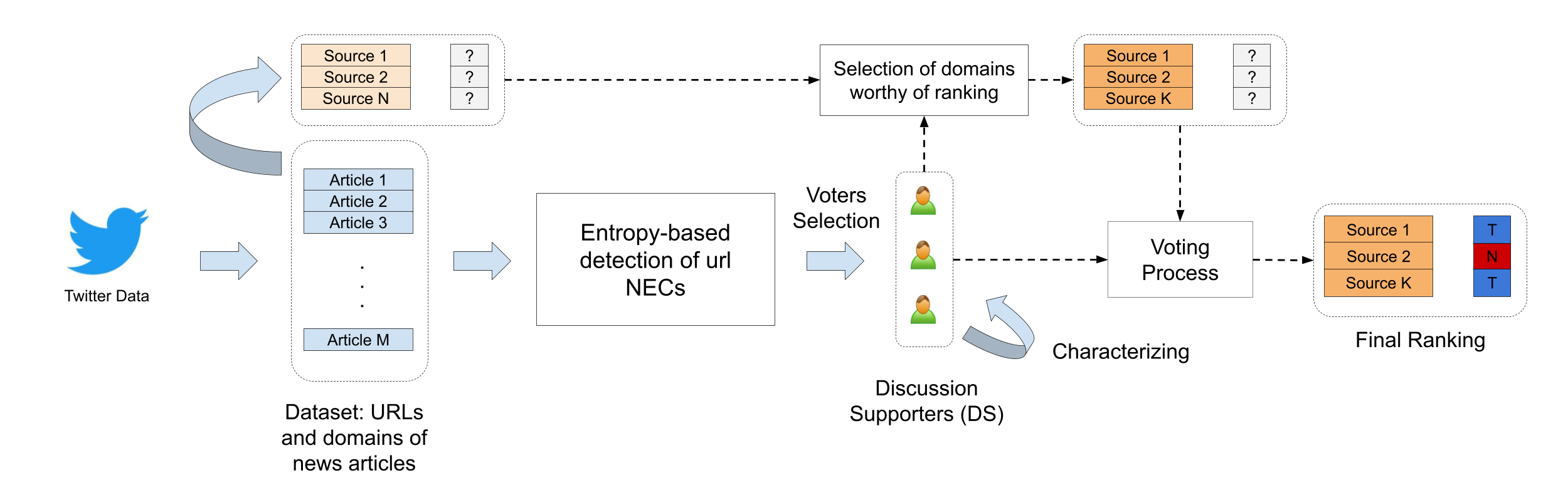}
        \caption{Schematization of the procedure for classifying the online publisher trustworthiness\label{fig:methodology}}
        \Description{Schematization of the procedure for classifying the online publisher trustworthiness}
    \end{figure*}

Our proposal consists of four steps, depicted  in Figure~\ref{fig:methodology}.
In the first step, that of data collection, we consider the posts published on the social network of interest regarding a specific narrative - in our case the social network is Twitter/X and the narrative is about the Covid-19 vaccination campaign - and we extract those URLs in the posts that point to online news. In parallel, we extract the domains related to these URLs, that is, the list of media outlets that published the online news.

The second step is to analyze the links between the users and the URLs in their posts. 
As in~\cite{GuarinoPGC21,DBLP:journals/corr/abs-2308-01750}, we use a bipartite network of users and URLs. We compare the real-world observations to an entropy-based benchmark: 
%
If two URLs appear in the posts of the same users significantly more often than the benchmark, we conclude that these two URLs cause users to share the same information diet in a statistically significant way, and we project an association between the two URLs.
To observe and characterize the network of URLs obtained in this manner, we employ a community detection algorithm. The resultant clusters of URLs are referred to as \emph{news engagement communities}, NECs for brevity.
As pioneered in previous work~\cite{mattei22,GuarinoPGC21}, the validated projection has the property of bringing out communities of URLs belonging to domains that are almost homogeneous from the point of view of trustworthiness. 
Furthermore, the URLs that pass the statistical test and thus become part of the URLs NEC are usually a much smaller number than the initial number in the original dataset, so their nature can be studied rather easily, for example by resorting to tags associated with URLs domains through organizations such as NewsGuard. 

The third step is to collect all users who have shared at least one URL belonging to a URL NEC on the social network. These users are called {\it Discussion Supporters}.
These users have shown interest in news stories, which in turn have contributed significantly to the formation of a shared discussion.
Here, we want to use {\it Discussion Supporters} to estimate the trustworthiness of all online publishers they support via a classification procedure (the fourth and last step).  

\textbf{What we do in this paper:} We present a novel approach for classifying the trustworthiness of online publishers, 
based solely on observed interactions between users and URLs in data collected from active social media discussions. Our work considers Twitter/X as a reference platform.
Our proposal has three main pillars. The first, using external sources to tag a knowledge base: we will rely on NewsGuard tags to label the domains to which URLs in communities belong.  Second, we use an entropy-based null model to identify the URLs that participate in communities and the Discussion Supporters.
Third, use the Discussion Supporters 
to classify the trustworthiness of unclassified publishers.\\

\textbf{What research questions we want to answer:} 
\begin{itemize}
    \item \textbf{RQ1:} Can we use the social interactions between online news publishers and online news consumers to assign a trustworthiness label to the publishers?;
    \item \textbf{RQ2:} If the answer to RQ1 is yes, how many of the publishers with whom consumers interact can we assign a trust label to? (In other words, does the proposed methodology provide sufficient coverage?);
    \item \textbf{RQ3:} If the answer to RQ1 is yes, what is the classification performance? 
\end{itemize}

\textbf{Contributions:}

\begin{itemize}
    \item \textbf{Automatic Classification of Publishers' Trustworthiness:}
    We present a novel method for automatically classifying the trustworthiness of online publishers, exploiting the interactions within social media discussions between two central actors: publishers (news producers) and users (news consumers). The method is designed to work effectively in real-world applications, where the delicate balance between classification performance and cost is crucial. 
    
    \item \textbf{Recognition of Influential Users Supporting Online News Flow:}
   We contribute by identifying users within social discussions who play a critical role in supporting the flow of online news articles. These users, called Discussion Supporters, are particularly active in sharing common narratives and publishers. This identification not only improves our understanding of user behavior, but also helps to formalize the publisher classification problem in the context of ongoing social media discussions.
   
    \item \textbf{Identification of Worthy-to-be-ranked Domains:}
    Our work includes the identification of a list of \textit{worth-to-be-ranked domains}. This curated list serves as a valuable guide for directing traditional annotation processes towards unknown publishers that may be relevant for classification, and thus may be useful for organizations such as NewsGuard, MediaBias Fact Check, the Global Disinformation Index, etc.
\end{itemize}


\section{Useful definitions}\label{sec:problem_def}
Here, we give some definitions that will be useful throughout the paper.
Let be \(A\) a set of online news articles whose links are shared in a specific social media discussion, with $n=|A|$. 
 

Let be 
$P$ a set of online news publishers, with $m=|P|$. 
Each publisher \(p\) is assigned a \(trust_{p}\) score, an integer in the range \([0, 100]\), indicating its trustworthiness. From these scores, we define a set \(L\) of distinct levels of trustworthiness associated with publishers,
where \(q=|L|\) is the number of trustworthiness levels. The set \(P_{l}\) contains the publishers associated with the trustworthiness level  \(l\).
 
 
Each article \(a\) is associated with a single trustworthiness level \(l\), determined based on the \(trust_{p}\) score inherited from the publisher $p$ of the article.

In this work, we will leverage social interactions between online users and publishers to characterize the trustworthiness level of the latter. We will follow different strategies, by changing the set of users employed in this procedure. We will call this set {\it voters} and denote it as \(V\), disregarding the adopted strategy. 


\textbf{Publisher coverage:} Given the sets \(V\), the \textbf{publisher coverage} \(PC_V\) of \(V\) is defined as 
the quantity of publishers posted by users in \(V\).



\textbf{Publisher coverage wrt \(l\):} Given the trustworthiness level \(l\) and the sets \(V\) and \(P_{l} \subseteq P\), the \textbf{publisher coverage wrt \(l\)} of \(V\), called \(PC^{l}_V\),  is defined as the quantity of publishers posted by users in \(V\) also contained in \(P_{l}\).


\textbf{Classification task:} 
We will develop a classifier for online publishers, named \(C_{trust}\), which associates the level of trustworthiness \(l \in L\) for each publisher 
\(p\) posted by the voters in $V$.  \(C_{trust}\) will leverage the characteristics of voters in  \(V\).

\noindent Since the classification will look at voters and their characteristics, we will need to reach two intermediate milestones before proceeding with the evaluation of publishers:

\textbf{How to select the voters:} Given \(P\), we need to select the set \(V\) such that the articles published by the voters in $V$ ensure the maximum publisher coverage on \(P\).
The result of this task is the \textit{list of publishers worthy of ranking}.

\textbf{How to characterize the voters:} Given \(V\), we need to characterize each voter \(v\ \in V\) to extract relevant information for the classification task.




\section{Methods}\label{sec:methods}
In this section, we describe the dataset used for our analysis, explain how we approached the problem of selecting and characterizing voters, and implement the classification task.

\subsection{Data collection}\label{sec:data}
    
\begin{table*}[ht]
\begin{center}
\caption{Keywords used to collect tweets related to the Twitter discourse on the Covid-19 vaccination campaign. The keywords were searched in Italian, and their English equivalents are given on the right}
\label{tab:keywords}
    \begin{tabular}{l|l}
        \toprule
       \bf{Keywords} & \bf{English meaning}\\
        \midrule
        vax, vaccino, vaccini, vaccinarsi, novax &variants of the word `vaccination' \\&(`novax': an individual who is )\\&(against vaccination)\\
        Astrazeneca, Pfizer-BioNTech, Moderna, Sputnik& common Italian names of Covid-19 vaccines\\
        greenpass&the certificate guaranteeing that the vaccine has been taken\\& or that the disease has been cured\\
        \bottomrule
    \end{tabular}
\end{center}
\end{table*}
Our dataset consists of approximately 1.87 million tweets in Italian written by 136,000 users, with nearly 220,000 tweets containing URLs. The data were collected using the Twitter's streaming API from September 1 to September 24, 2021.

The data collection process was keyword-based and focused on the online discourse surrounding COVID-19 vaccination, see Table~\ref{tab:keywords}. 
 Twitter's streaming API captures every tweet that contains the specified terms in both the text of the tweet and its metadata.
It is not necessary to include every permutation of a given keyword in the tracking list. For example, the keyword COVID would include tweets that contain both COVID19 and COVID-19.

\textbf{Knowledge base}
To build our knowledge base, we inherit the ratings and labels that NewsGuard associates with news sites\footnote{\url{https://www.newsguardtech.com/ratings/rating-process-criteria/}}.

\begin{table}[ht]
    \begin{center}
    \caption{Domain tagging tags inherited from NewsGuard. The UNC tag indicates that NewsGuard has not yet tagged this domain}
    \label{table:domains-tags}
    \begin{tabular}{c|l}
        \toprule
        {\bf Label} & \textbf{Description}\\
        \midrule
        T & Trustworthy news publisher \\
        N & Untrustworthy news publisher \\
        UNC & Unclassified publisher\\
        \bottomrule
    \end{tabular}
    \end{center}
\end{table}

Table~\ref{table:domains-tags} provides an overview of the tags associated with news publishers. Based on a series of analyses performed by NewsGuard on the specific online news outlet, it is assigned a score between 0 and 100: if the score is greater than or equal to 60, the tag is T; otherwise, it is N. 
Processing tweets in our dataset, we find a total of $5,749$ news domains, of which $381$ are tagged as T, $116$ as N and $5,252$ as UNC (unknown to NewsGuard).

\subsection{Selection of voters through URL NEC}\label{sez:detecting_ds}

The primary challenge we face is the voter selection problem. This task involves identifying an optimal set of voters \(V\) that can effectively contribute to the evaluation of relevant publishers in the social media discussion under consideration, with a particular focus on identifying potentially low-quality publishers.

In this study, relevant publishers are those entities capable of creating compelling narratives that uniquely engage audiences and foster the attention of like-minded groups. In line with this definition, 
we select as voters those social users who 
support 
articles that have generated increased engagement among groups of people within the broader discussion.

For the following analyses, the concept of URL News Engagement Communities, introduced in \cite{DBLP:journals/corr/abs-2308-01750} and known as URL NECs, is crucial. The main idea is to infer similarities between URLs based on the way they are shared on the platform by different users. In practice, we are interested in all pairs of URLs that are shared more frequently than would be expected based on their frequency alone.
We model users sharing URLs on the social platform as a bipartite network, wherein URLs and users respectively constitute the two distinct sets of nodes (or layers, in network terminology), and a link indicates that the user is sharing the URL. Then, for each pair of URLs, the number of co-occurrences of users is counted: these represent the users who shared both URLs. Finally, these co-occurrences are compared to a statistical benchmark to determine their significance. In the present case, we use a maximum entropy null model, since it provides an unbiased benchmark for the analysis~\cite{Saracco2017,Cimini2019}.

In a nutshell, the idea, borrowed from statistical physics~\cite{Jaynes1957}, consists of 3 steps: first, starting from the real network, define a set of graphs -called \emph{ensemble}-, all having the same number of nodes but representing all possible link configurations, from the empty network to the fully connected one. Then, each representative of the ensemble is assigned a different probability by a constrained maximization of the entropy associated with the ensemble. In this step, the constraints have to be chosen carefully: in order to highlight deviations from a standard behavior, the constraints should represent fundamental properties for the description of the real system. In many cases, the degree sequence, i.e. the number of connections for each node, is particularly informative. When the constraints are the degree sequences of both layers of the bipartite network, the null model is called \emph{Bipartite Configuration Model} (or BiCM~\cite{Saracco2015}).

For each pair of URLs, we count the number of co-occurrences in the real networks, i.e. the number of users who posted both URLs, and assign a p-value to each co-occurrence by comparing the observed value with the relative BiCM-induced distribution. 
P-values are then statistically validated (in this case we choose the significant level $\alpha=0.05$). To perform a multiple testing hypothesis (we have as many p-values as the number of possible pairs of URLs), we implement FDR (\emph{False Discovery Rate}), since it controls the proportion of false positives~\cite{Benjamini1995}.

The result of this procedure is a validated monopartite network of URLs, in which two nodes are connected if they were significantly shared by the same accounts. 

In order to characterize similar URLs, i.e. those that appear to be more connected and form communities within the resulting monopartite network, we apply a community detection algorithm. To detect communities, we choose a well-established algorithm, namely Louvain~\cite{Blondel2008}. It is important to emphasize that, while there are a few community detection algorithms for bipartite networks - for example, extending the concept of modularity to bipartite networks \cite{Barber2007,Guimera2007} - we choose a monopartite community detection algorithm on the validated network because such a procedure already ensures that only the significant signal is analyzed, cutting out all uninformative fluctuations.

The resulting partition was named in~\cite{DBLP:conf/icwsm/PratelliP22} \emph{News Engagement Communities} of URLs (URLs NECs for short), i.e. URLs grouped together with respect to the interest that users have shown in them.


\begin{table}[ht]
\begin{center}
\caption{Statistics for URL NECs}
\label{tab:url_nec_details}
\begin{tabular}{lcccc}
\toprule
ID &  No. users &   Distinct URLs &  No. publishers &   No. URLs \\
\midrule
          4 &   7422 &                  223 &       71 &  38731 \\
         11 &   1064 &                    21 &        4 &   1681 \\
          6 &    876 &                    87 &        5 &   2019 \\
          1 &    674 &                    79 &        9 &   2234 \\
          9 &    584 &                    58 &        1 &   1238 \\
          7 &    521 &                   64 &        6 &   1613 \\
          5 &    311 &                    23 &        6 &    562 \\
         13 &    253 &                     3 &        3 &    304 \\
         12 &    175 &                  28 &        4 &    408 \\
          0 &    161 &                    55 &        1 &    365 \\
          8 &    149 &                   25 &        1 &    308 \\
          3 &    101 &                   79 &        1 &    393 \\
         10 &     65 &                   27 &        9 &   1557 \\
         14 &     42 &                  4 &        2 &     59 \\
          2 &     23 &                  3 &        3 &     32 \\
\bottomrule
\end{tabular}
\end{center}
\end{table}

\begin{figure}[h]
    \centering
    \includegraphics[width=.99\linewidth]{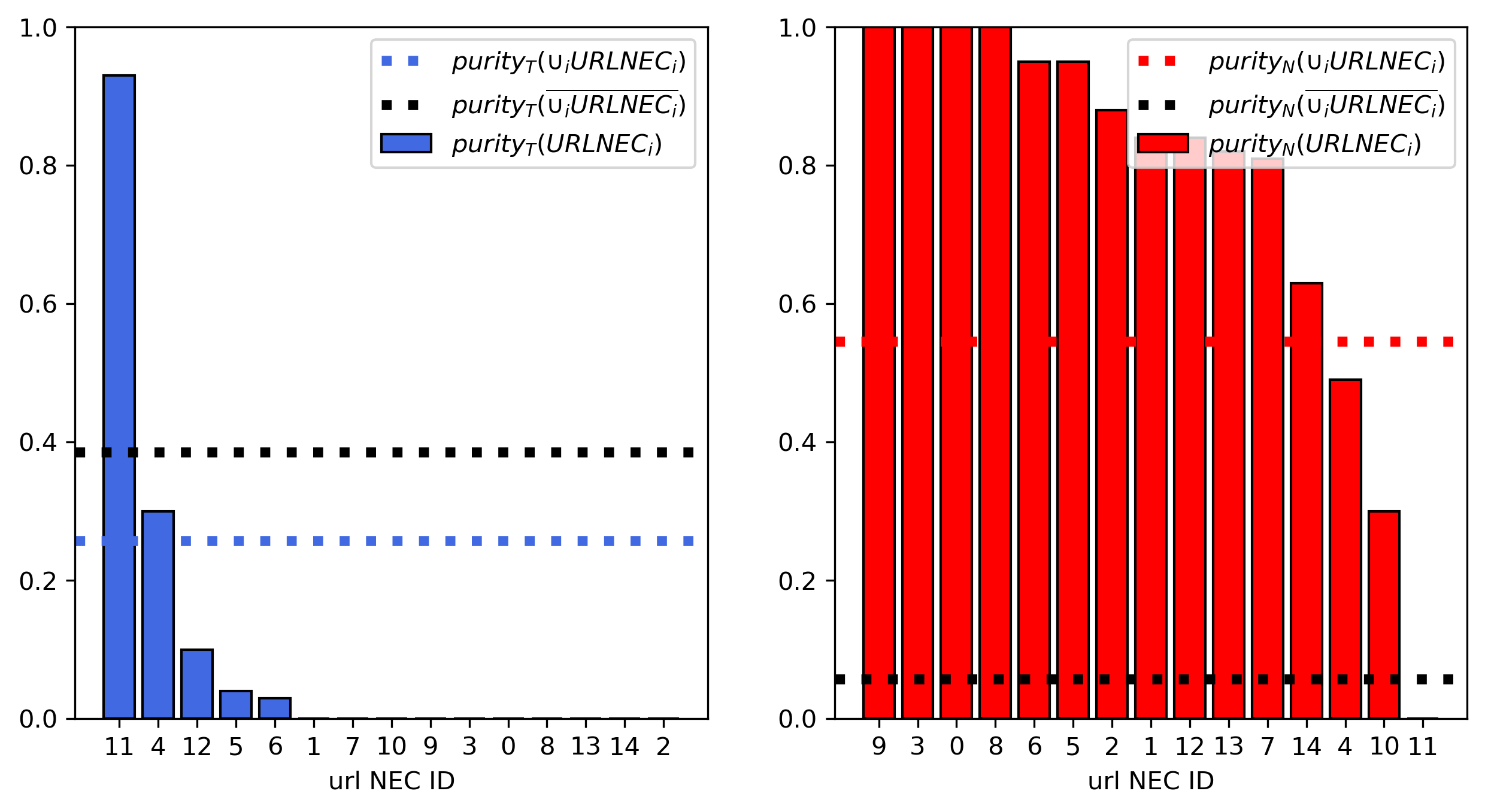}
    \caption{Purity levels of URL NECs. Trusted URLs on the left, untrusted URLs on the right\label{fig:vaccine_homog}}
    \Description{}
\end{figure}

We have 51,504 URLs that fall into URL NECs, about 22\% of the total. Table~\ref{tab:url_nec_details} shows some data about these URL communities. ID is the individual community identifier, No. users is the number of users who have shared URLs in the community, No. publishers is the number of online media outlets that have published the URLs.

To characterize and observe the properties of emergent URL NECs, we analyze the homogeneity of a URL's community in terms of publisher trustworthiness. To perform this task, we use a metric called `purity level' (see the Appendix for a detailed definition). 
This metric expresses how often trustworthy (and untrustworthy) publishers are associated with URLs in a URL NEC. 

Figure~\ref{fig:vaccine_homog} 
shows the purity level of each community in terms of links T (trustworthy) and N (untrustworthy). 
As noted in previous work~\cite{GuarinoPGC21,mattei22}, the URL validated projection
has the property of producing communities of URLs belonging to
publishers that are almost homogeneous from the point of view of
(un)trustworthiness.

In summary, URL NECs tend to capture news articles that (i) are shared by a common audience (similar like-minded people, as evidenced by their statistically significant support for the same news), and (ii) have a homogeneous level of trustworthiness defined at the publisher level. 
It is important to emphasize that similar results, specifically homogeneous NECs in terms of publisher trustworthiness levels, can also be obtained using other community detection algorithms, such as InfoMap \cite{rosvall2008maps}.
Since each URL considered in the present analysis is linked to a news article in a one-to-one correspondence, and since we are ultimately interested in using the news articles to infer the reliability of the users who share them, we will call \(A_{val}\) the validated monopartite network of URLs/news articles obtained above. Note that $A_{val} \subseteq A$, defined in~\ref{sec:problem_def}: the validated articles of the URL NECs represent a subset of all articles in our dataset.

We will use the validated news articles in \(A_{val}\) to characterize voters. Specifically, we can think of selecting as voters those users who have engaged with validated articles at least once. We call these users \textit{Discussion Supporters} (DS): users who have played a central role in supporting the spread of the set of URLs that characterize the current online discussion. 
In the next section, we will better investigate how to characterize voters, whether \textit{Discussion Supporters} or other sets of users.

\subsection{Endogenous characterization of voters}\label{sez:voters_char} 
Explicitly modeling the endogenous preferences of users based solely on their social network information is a non-trivial task. Building on similar methods used in previous work~\cite{dou2021user}, we propose to use the messages posted by each user to characterize them. Specifically, our approach consists of approximating each voter's tendency to be a spreader (or non-spreader) of low-quality content. We measure this propensity using the average trustworthiness score of news articles shared by the user. To achieve this, we take a publisher-based approach and consider the tag assigned to the news publisher by expert annotators. Despite the approximation inherent in our measure, we argue that the use of expert annotations can ensure high quality ratings. 

In the characterization procedure, we do not consider quoted tweets. We exclude them from the analysis because we cannot know for sure whether they are supportive or critical of the original tweet, since we do not address the content added by the user.

We propose four different strategies for characterizing voters, depending on whether and how we adopt the URL NECs.

\begin{itemize}
    \item \textsc{DS-URL-NEC}: According to this first  strategy, the voters are the Discussion Supporters, \(V = DS\). We characterize the Discussion Supporters by considering only those articles that {\it they have shared on the social network in a specific discussion and that are part of at least one URL NEC}. We denote these articles as \(A_{v_i}|_\textsc{DS-URL-NEC}\).   
    Following this first strategy, the URL NECs are used both to detect the set of voters \(V = DS\) and to identify the subset of news articles for their characterization. 
    \item \textsc{DS-ALL}: According to this second strategy, the voters are the Discussion Supporters, \(V = DS\). However, for their characterization, we consider {\it all the news articles that the voters have shared in the social network in a given discussion}.  
    We denote these articles as \(A_{v_i}|_\textsc{DS-ALL}\).   
    Thus, in this case, the URL NECs are only used to detect the set of voters \(V\).  
    \item \textsc{DS-ALL-WO-USR-NEC}: According to this third strategy, the voters are the total set of users \(U\) minus the Discussion Supporters \(V = U - DS\). Each voter \(v_i\) is characterized by considering all the news articles \(A_{v_i}|_{DS-ALL-WO-USR-NEC}\) shared by the voters on the social network in a given discussion.
    \item \textsc{USERS-ALL}: This strategy does not consider the notion of URL NECS.  Here, the voters are the entire set of users and the news articles are those shared in the discussion:  \(A_{v_i}|_{USERS-ALL}\).
        
\end{itemize}

For all strategies, we characterize the articles shared by voters by the average of the trustworthiness scores NewsGuard assigned to the article publishers. As an example, consider the articles that each voter has shared in the social network using one of the strategies described above. We assume that the \(i-th\) voter has shared 10 news articles, 5 from publisher \(x\) and 5 from publisher \(y\). We also assume that according to our external source NewsGuard, publisher \(x\) has a trust score of =60 and publisher \(y\) has a trust score of =90. Then the \(i-th\) voter is assigned a characterization value equal to the arithmetic mean \((60*5 + 90*5) / 10 =75\).

We also consider an additional characterization of voters by considering the variability of their `information diet,' i.e., how many different publishers a voter has shared in the social network. The goal of this additional characterization is to evaluate whether variations in voters' information diets can influence the reach and performance of our proposal.

\begin{figure}[h]
    \centering
    \includegraphics[width=\linewidth]{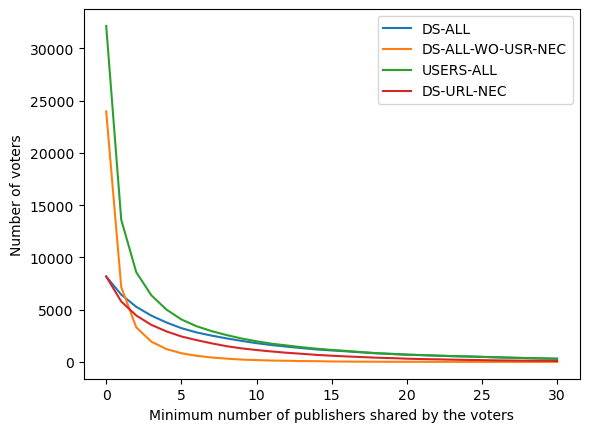}
    \caption{Number of voters {\it wrt} the minimum number of shared publishers and the adopted strategy \label{fig:voters}}
\end{figure}

Figure~\ref{fig:voters} shows the number of voters {\it wrt} the chosen strategy for their selection and the minimum number of shared publishers. 

\subsection{Classification of online publishers via voters}\label{sec:classification}
Here we propose an implementation for the classification task defined in Section~\ref{sec:problem_def}.
We consider two levels of trustworthiness, such that \(L=\{T, N\}, (|L|=2)\). The label \textsc{T} corresponds to trustworthy publishers, while \textsc{N} denotes untrustworthy publishers. These labels correspond to those used by NewsGuard (see Section~\ref{sec:data}). The classification task is binary.

To determine whether a publisher posted by the voters in $V$ belongs to the \textit{T} or \textit{N} class, we take the characterizing values of all voters who have interacted with articles from that publisher, and again perform an arithmetic mean. For example, if we have 10 voters who have interacted with publisher \(p\), 5 with characterizing value =75 and 5 with characterizing value =60, it simply follows that publisher \(p\) is associated with a value of \(75*5 + 60*5 /10 =67.5\). We can then proceed with the implementation of a simple decision tree model with a maximum depth of one, due to its simplicity and interoperability. What we will show as the result in the following sections comes from a classifier to which we give as input, for each publisher, its score (e.g., 67.5) as the only feature and the label available from Newsguard. We use the scikit-learn Python library, which provides implementations for (i) the decision tree and (ii) a 10-fold stratified cross-validation (preserving the percentage of samples for each class in the folds).

 In the following, we will present the classification results for those publishers annotated by NewsGuard. Obviously, unclassified publishers cannot be used to evaluate our proposal, as we lack ground truth information for them.

\section{Results}\label{sec:results}
This section presents the results of our analysis, first on the ability to assign trustworthiness labels to the largest number of publishers in the initial dataset, and then on the classification performance.

\begin{figure}[h]
    \centering
    \includegraphics[width=.8\columnwidth]{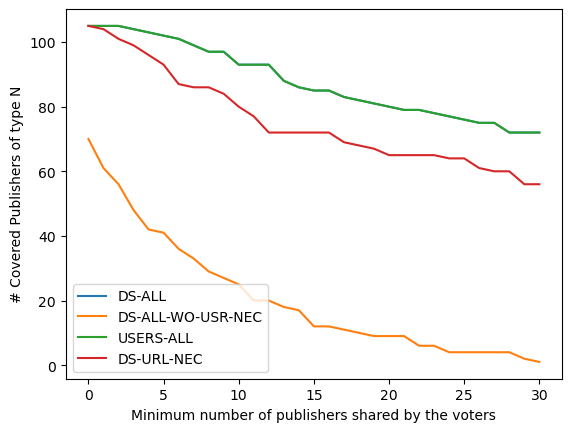}
    \includegraphics[width=.8\columnwidth]{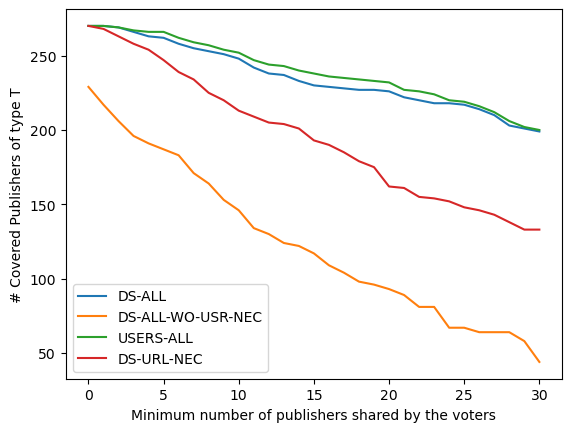}
    \caption{How many publishers can we reach given the original number of publishers, the set of voters, and the minimum number of publishers shared by the voters?}
    \label{fig:sources_coverage}
\end{figure}

\subsection{Coverage of Publishers}\label{sec:res_coverage}


\begin{table}[ht]
    \begin{center}
    \caption{Percentage of publisher coverage}
    \label{tab:traffic_source_coverage}
    \begin{tabular}{lrrr}
        \toprule
        Strategy & T & N & UNC \\
        \midrule
        DS-ALL / DS-URL-NEC &  70.87 &  90.52 &  29.74\\
        \bottomrule
    \end{tabular}
    \end{center}
\end{table}

In Table~\ref{tab:traffic_source_coverage}, we show the publisher coverage \(PC^{l}_V\) when the voters in \(V\) are the {\it Discussion Supporters} and the minimum number of publishers shared by the voters is 0. Given this configuration, we note that both DS-ALL and DS-URL-NEC contribute equally to the coverage calculation, since both consider the same set of voters. 
%
We consider the trust levels $L =\{$T, N, UNC$\}$. 
We capture $\sim71\%$ of the trusted publishers. 
%
We are also able to reach \(90.52\%\) untrustworthy publishers. 
 Thus, we can say that, especially for publishers of type N,  the coverage is particularly large.
 
 The coverage also includes \( \sim 30\%\) of unclassified publishers\footnote{Unclassified publishers are those publishers for which we do not have a trustworthiness label assigned by an external source (NewsGuard in this study).}. In terms of helping organizations like NewsGuard identify news media for analysis, these UNC publishers are promising because they are publishers shared by voters who were also interested in peculiar news (i.e., news whose links end in URL NECs).

In Figure~\ref{fig:sources_coverage}, we show how many T and N publishers we can reach, considering all the strategies defined in Section~\ref{sez:voters_char} and the minimum number of publishers a voter must engage with to be considered. 
On the y-axes, the top panel shows the number of N-type publishers reached, and the bottom panel shows the number of T-type publishers reached (in our case study, we have a total of 116 N-type publishers and 381 T-type publishers, see Section~\ref{sec:data}). 

In the top panel, we can see that strategies such as \textsc{DS-ALL}, \textsc{DS-URL-NEC}, and \textsc{USERS-ALL} consistently provide the best coverage, even as the threshold for the minimum number of publishers increases. On the contrary, the \textsc{DS-ALL-WO-USR-NEC} strategy shows limitations, providing the least coverage for untrusted publishers. This underscores the effectiveness of relying on URL NECs to construct an efficient list of publishers worth annotating in terms of coverage. 
In the bottom panel, we observe a similar pattern for the coverage of trusted publishers.

\subsection{Classification Performance}

\begin{figure}[h]
    \centering
    \includegraphics[width=\linewidth]{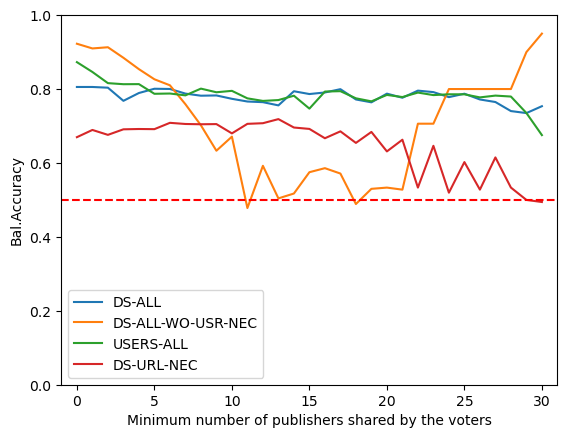}
    \caption{Publisher classification performance\label{fig:performances}}
\end{figure}

\begin{figure}[h]
    \centering
    \includegraphics[width=\linewidth]{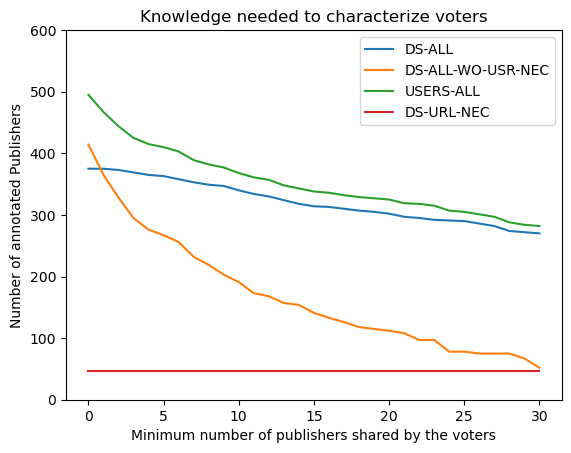}
    \caption{Initial knowledge in terms of publishers annotation  \label{fig:knowledge}}
    \Description{XXX.}
\end{figure}

Figure~\ref{fig:performances} shows the classification results obtained on the covered publishers, using a simple decision tree (Section \ref{sec:classification}). 
The baseline, set at $0.5$, serves as a reference point.

Quite intuitively, if the voters are all the users in the initial dataset, and their characterization is calculated from all the news they have shared (strategy \textsc{USER-ALL}), the classification results are the best, with a peak in Balanced Accuracy at $0.806$.

Let us now analyze what happens when we consider strategies that consider \(V =DS\), namely \textsc{DS-ALL} and \textsc{DS-URL-NEC}. Both strategies outperform the baseline. 
Noteworthy is the fact that the strategy where we characterize Discussion Supporters with all their shared URLs, \textsc{DS-ALL} produces results comparable to the \textsc{USER-ALL} strategy. 
As for the strategy that considers only the ground truth of the publishers in the URL NECs, i.e., \textsc{DS-URL-NEC}, it leads to worse results than the  \textsc{DS-ALL} strategy, but still achieves peaks of Balanced Accuracy = $0.719$, and at a lower cost in terms of initial labeling, see Section~\ref{sez:voters_char}.

Finally, considering the classification of publishers only on the basis of users who did not share anything in the URL NECs, we underline that these users cover a very small part of the T and N URLs in the original dataset of URLs, as can be seen in Figure~\ref{fig:sources_coverage}. Thus, while it is true that in some cases, depending on the number of shared publishers, the accuracy is comparable to the other strategies,  it is also true that these users are not useful for our purposes, given the small coverage.

\section{Discussion}\label{sec:discussion}
In our quest for the optimal classification solution, we must make a trade-off between three crucial aspects: publisher coverage, classification accuracy, and the cost of acquiring the knowledge required to characterize voters who contribute to publisher classification.

\textbf{Knowledge vs. Cost.} Figure~\ref{fig:knowledge} shows the number of annotated publishers required to characterize voters and thus classify additional publishers.
Regardless of the minimum number of publishers shared by the voters, if we follow the strategy \textsc{DS-URL-NEC}, the knowledge is constant, equal to the number of publishers shared by DS whose URLs are also part of a URL-NEC. In the specific case of our case study, we need the trust labels of 46 publishers. If we follow the \textsc{USER-ALL} and \textsc{DS-ALL} strategies, we need the annotation of at least 300 publishers (in case the voters have a rather diverse information diet, at least 25 shared publishers per voter). 
Since the manual process required to annotate even a single publisher is time-consuming and tedious, if we can afford a classifier with lower accuracy (see Figure~\ref{fig:performances}), \textsc{DS-URL-NEC} is obviously preferable.

\textbf{Accuracy vs. Cost.} Still referring to Figure~\ref{fig:performances}, the strategies \textsc{USER-ALL} and \textsc{DS-ALL}  give the best results in terms of balanced accuracy. However, as shown in Table \ref{fig:knowledge}, both strategies require a considerable amount of initial knowledge compared to \textsc{DS-URL-NEC}. Although \textsc{DS-URL-NEC} achieves a Balanced Accuracy below \textsc{USER-ALL} and \textsc{DS-ALL}, it outperforms the baseline (see Table \ref{fig:performances}) and can provide acceptable performances; we remember that for the (cheaper) \textsc{DS-URL-NEC} strategy we reach a peak of $0.719$, while with the most expensive one (i.e.,  \textsc{DS-ALL}) we achieve $0.806$.

\textbf{Coverage vs. Knowledge.} In real-world scenarios, prioritizing a clear list of relevant publishers for classification is challenging, leading to potential gaps in coverage and the risk of losing the evaluation of low-credible publishers. To address this challenge and ensure coverage of low-credibility publishers, we propose to generate a list of publishers worth annotating using an entropy-based procedure (see Section~\ref{sez:detecting_ds} for details). In Section~\ref{sec:res_coverage}, we observed that two strategies proposed in this work, namely \textsc{DS-ALL} and \textsc{DS-URL-NEC}, provided the best publisher coverage with different trade-offs in terms of knowledge.

\textbf{Twitter's Change of Ownership and the Rise of X.}
In late October 2022, the American social media company Twitter, Inc. underwent a significant transformation as it came under the ownership of Elon Musk\footnote{\url{https://www.nytimes.com/2022/10/27/technology/elon-musk-twitter-deal-complete.html}}. This transition ushered in a series of radical changes and reforms that included both managerial and technical aspects.

A key development for the scientific community was the elimination of Twitter's free API tier by February 2023, to be replaced by a `basic paid tier'\footnote{\url{https://twitter.com/XDevelopers/status/1621026986784337922}}. For researchers and developers, this shift meant that Twitter content would no longer be available for research purposes without subscribing to a significantly different paid plan. It also posed a challenge in terms of re-hydrating the datasets currently in use.

Although our dataset was collected during a period of free access (from September 1 to September 21, 2021), the policy appears to be unchanged at the time of writing this manuscript\footnote{\url{https://developer.twitter.com/en/docs/twitter-api/getting-started/about-twitter-api}, consulted February 27, 2024.}. Thus, we acknowledge the potential obstacles to the reproducibility of the experiments presented here.

However, we maintain that our methodology remains highly adaptable to other online social networks. Indeed, it relies on a core principle: the analysis of account activity related to the sharing of news publishers' URLs. We argue that extending this approach to alternative social platforms is feasible.

\section{Related Work}
Assessing the trustworthiness of a publisher, a key factor in distinguishing real from false news~\cite{bazmi2023multi}, is of great importance in understanding the spread of misinformation online~\cite{lazerScience2018}. Social platforms use trustworthiness scores to control user exposure to content from unreliable sources and highlight credible sources of information~\cite{Nadarevic2020}.

{\bf Media Communication Experts.} Various journalistic organizations and indexes, including entities such as NewsGuard, MediaBias Fact Check, and Ad Fontes Media, engage in studies to evaluate the transparency and trustworthiness of online news publishers. 
While these organizations use different criteria for evaluation, recent research has revealed a remarkable convergence in the labels that each entity assigns to individual media outlets. This convergence underscores the consistency and reliability of these organizations' assessments of media trustworthiness~\cite{PNASPennycook2023,DBLP:conf/icwsm/PratelliP22}. 

Unfortunately, this evaluation is a laborious process, especially in terms of time. The Global Disinformation Index, for example, uses a meticulous process of selecting expert annotators who specialize in a country's online information system. After a training period, the annotators curate a selection of online newspapers that aptly represent the country's information landscape. They then conduct a manual analysis of the sites, delving into details such as newspaper ownership and funding sources.
This initial phase is followed by a thorough content analysis of a sample of articles from each outlet to identify unreliable, sensational, and/or propagandistic content. 
The whole process can take up to several months, and the annotators must of course also go through a `rethinking' phase if individual assessments differ greatly from one another~\footnote{The first and third authors are familiar with the process, having been involved as annotators in the
country study of the Italian landscape.
}

{\bf Article classification.} To better support and streamline the evaluation process, recent work has proposed automatic methods for estimating the publisher's trustworthiness based on the individual classification of news articles.  The study in~\cite{bohavcek2022fine} considers four levels of trustworthiness and associates one of them with news articles in Czech. 
The work by Przybyla~\cite{DBLP:conf/aaai/Przybyla20} chooses a binary classification (trustworthy vs. untrustworthy publisher) and achieves an average accuracy of $0.878$ by classifying at the level of the news article. 
Recent research~\cite{MenczerGPT2023} explores the use of Chat-GPT's API to rank online news publishers and achieves an area under the curve of 0.89 in a binary ranking scenario (trustworthy vs. untrustworthy). 

{\bf Social interactions.} In this paper, we aim to predict the level of trustworthiness of a publisher not from the analysis of the published content, as done in previous literature, but rather from the interactions that the publisher has with the users of a social network. Specifically, by interactions we mean the posting and reposting by social network users of links to news published by the publisher. 

In the past, the literature has well demonstrated how studying the dynamics of information dissemination in the social network is critical to uncovering false and/or biased content and the actors acting to disseminate that content~\cite{,shao2018anatomy,PENNYCOOK2021388,DBLP:journals/corr/abs-2304-02983,DBLP:journals/cacm/FerraraVDMF16,DBLP:journals/epjds/CaldarelliNPPS21}. The spread of information is very different whether it is reliable news or not, for example, fake news spreads faster and more widely than the truth on social media~\cite{vosoughi2018spread}. 

Some recent work, such as that in~\cite{gravino2022supply, shu2019beyond, su2023hy, dou2021user, shrestha2023joint}, has explored the social relationships between 
the social relationships between content producers and readers to reveal hidden information dynamics. 
A paper that has points in common with ours, though not intended to rank the degree of trustworthiness of a publisher, is that of Gravino et al.~\cite{gravino2022supply}: the authors analyze the relationships between online news supply and online news demand. 
This study of the interplay is essential to reveal the mechanisms of information dynamics. 
To instantiate demand, the authors consider user interests as reported by Google Trends.
To instantiate the supply, they collect news articles from various online media, tagged as both "questionable" and "all".  
A very interesting result when comparing demand and supply is that users' interests are closer to content published by untrustworthy publishers than by mainstream publishers. Or, the other way around, that untrustworthy publishers are more likely to satisfy readers' interests. Most likely, analyzing readers' interests is one of the mechanisms that untrustworthy publishers use to attract users. Our work exploits exactly this mechanism, i.e., engaging readers by satisfying their interests. URL NECs are exactly the effect of this mechanism, groups of URLs, moreover homogeneous from the publisher's point of view, that have attracted the attention of a statistically significant number of users. 

Other attempts to detect low-credibility content focus on exploiting the user-news relationship, modeled as a graph~\cite{shu2019beyond, su2023hy, dou2021user, shrestha2023joint, su2023hy}.

In~\cite{shu2019beyond}, the authors address the novel problem of exploiting social context for fake news detection and propose a tri-relationship embedding framework, TriFN, which simultaneously models publisher-news relationships and user-news interactions for fake news classification.
The work in~\cite{dou2021user} introduces a preference-aware approach to fake news detection. Given a news article and its engaged users on social media, the method extracts exogenous context as a news propagation graph and encodes endogenous information based on users' historical posts and news texts. Exogenous and endogenous information are fused using a GNN encoder. The final news embedding, consisting of the user engagement embedding, is fed into the neural classifier to predict news credibility. While this approach works at the news level, our work focuses on the publisher level. 
Another recent work~\cite{shrestha2023joint} proposes a method that uses Relation Graph Convolution Networks to jointly predict the credibility of news, users, and news publishers in the news ecosystem. They model the news ecosystem as a heterogeneous graph composed of different types of nodes, such as news, users, and publishers, where links exist between different types of nodes. A unique node representation is created for each node type.

To the best of our knowledge, our work represents the first application of an approach based on complex network analysis to predict the trustworthiness of a news publisher. The key point is the extraction of relevant information, namely publishers and voters, from an ongoing social media discussion. This extraction is enabled by the detection of emerging URL NECs. 

\section{Conclusions}
This study presents a novel method for automatically classifying the trustworthiness of publishers, starting from limited initial knowledge. Our method exploits social media interactions between two key actors: publishers (news producers) and social users (consumers). The primary goals of our approach are twofold: 1) to identify a list of publishers worthy of annotation, and 2) to assemble a pool of voters capable of making informed judgments based on their historical behavior, thereby quantifying their tendency to spread trustworthy or untrustworthy news.

Our approach shows commendable performance even in scenarios with limited prior knowledge. It effectively navigates the tradeoff between effectiveness and cost, demonstrating its adaptability in contexts where extensive knowledge is limited. Furthermore, our observations highlight the method's ability to achieve robust coverage of publishers, including those classified as untrustworthy. These results underscore the potential of our approach in real-world applications where the trade-off between performance and cost is critical.


\textbf{Limitations:} Like other methodologies, our approach has some clear limitations. First, it depends on the context analyzed: in this case, the focus was on the acceptance of the COVID-19 vaccination in Italy and the news sources analyzed were the most active in covering this topic. Changing the topic would have resulted in a different data set, in which the interactions between the different accounts would have been different, and therefore the procedure would have yielded a different list of news sources to annotate. Such a point can be turned into a potentiality of our methodology: the present annotation strategy is topic-specific. In other words, 
our process identifies the most relevant news sources to annotate for the specific discussion. 

Including different online debates in the dataset would provide a finer description, where the relevance of different news sources is characterized at the topic level. 
However, it is important to emphasize that when different topics are present in the same dataset, multiple situations may occur. If the topics are somewhat disconnected from each other, such as the US presidential election and \textit{buccellato di Lucca}\footnote{\url{https://en.wikipedia.org/wiki/Buccellato_di_Lucca}}, we expect that our methodology will be less effective since the Italian receipt could break the ``bubble'' of US election propaganda and make the signal weaker. However, if the debates captured by the dataset correspond to arguments that are part of the same narrative - for example, migration from North Africa to Europe and COVID-19 vaccination, about which the Italian far-right conspiracy theorists are particularly active - then we expect the characterization of the cluster of domains to be finer. This topic will be the subject of further research\footnote{We are grateful to the reviewer who gave this suggestion.}.

Lastly, our proposal is currently limited to 
single social platform. Testing our procedure on 
different social platforms will be the subject of future research.


\begin{acks}
\small
Work partially supported by project SERICS (PE00000014) under the NRRP MUR program funded by the EU - NGEU; by Integrated Activity Project TOFFEe (TOols for Fighting FakEs) \url{https://toffee.imtlucca.it/}; by IIT-CNR funded project re-DESIRE (DissEmination of ScIentific REsults 2.0); by project “CODE – Coupling Opinion Dynamics with Epidemics”, funded under PNRR Mission 4 "Education and Research" - Component C2 - Investment 1.1 - Next Generation EU "Fund for National Research Program and Projects of Significant National Interest" PRIN 2022 NRRP, grant code P2022AKRZ9. 
\end{acks}

\bibliographystyle{ACM-Reference-Format}
\bibliography{sample-base}

\appendix

\section{Purity measure}
To analyze the homogeneity of URLs in a URL NEC in terms of publisher trustworthiness, we define a metric, called purity, which measures the frequency of URLs from reputable or non-reputable publishers in a URL NEC. 


Let be \(\text{URL NEC}_i\) the $i$-community and \(l_k\), k=1,2, the trustworthiness levels T and N.  We define $purity_{\small{l_k}}  (\text{URL NEC}_i)$
as the frequency of URLs belonging to $l_k$, i.e.:

\begin{equation}\label{eq:purity}
purity_{\small{l_k}} (\text{URL NEC}_i) = \frac{ | U_{i}^{\small{l_k}} | } { | U_{i} | },  
\end{equation}
where $U_{i} = \{URL_{1}, \dots, URL_{\small{n}} \}$ is the set of all the URLs in the $i$-community and $U_{i}^{\small{l_k}} \subseteq U_{i}$ is the subset of $U_{i}$ that contains only URLs with  trustwothiness level $l_k$. The purity defined in Eq.~\ref{eq:purity} can be interpreted as the probability of extracting an \(l_k\)-trustworthy URL from the $i$-th URL NEC.
If $m$ is the number of different URL NECs, we can define $purity_{\small{l_k}}(\cup_i\text{URL NEC}_i)$ as the frequency of URLs from $l_k$ domains in all URL NECs:
\begin{equation}\label{eq:overall_purity}
purity_{\small{l_k}}(\cup_i\text{URL NEC}_i)= \dfrac{\sum_{i=1}^m |U_i^{\small{l_k}}|}{\sum_{i=1}^m|U_i|}
\end{equation}

To have a benchmark for the purity of URL NECs, we also consider a purity measure for URLs that do not belong to any community: 
\begin{equation}\label{eq:unclust_purity}
purity_{\small{l_k}} (\overline{\cup_i\text{URL NEC}_i}) = \frac{ | U_{-1}^{\small{l_k}} | } { | U_{-1}| },  
\end{equation}
where the set of URLs that do not belong to any community is denoted as $U_{-1}$.\\

\section{Online Resources}
The X/Twitter IDs of the posts analyzed in this manuscript are available here (\url{https://doi.org/10.7910/DVN/6VDEBJ}), following the platform policy (\url{https://developer.twitter.com/en/developer-terms/more-on-restricted-use-cases}). At the time of writing, the X/Twitter policy on data sharing for academic research purposes is still unclear, since the platform APIs for downloading messages are not freely available, even for rehydrating tweet IDs (\url{https://developer.twitter.com/en/docs/twitter-api/getting-started/about-twitter-api}). Moreover, even if the application to access X / Twitter data is open to EU researchers (\url{https://developer.twitter.com/en/use-cases/do-research}), it is not clear what kind of resources will be available to researchers. 

The trustworthiness levels of news publishers come from NewsGuard, but there are restrictions on their availability as they are used under a NewsGuard license and are not publicly available. However, this data could be made available upon reasonable request and with permission from NewsGuard.

\end{document}